\documentclass[epj]{svjour}
\usepackage{graphics}
\usepackage{graphicx}
\usepackage{enumerate}
\begin{document}
\title{A Cautionary Tale: The Coulomb Modified ANC for the $\mathbf{1/2^+_2}$ State in $^\mathbf{17}$O} 
\author{N. Keeley\inst{1} \and K. W. Kemper\inst{2,3} \and K. Rusek\inst{3}
}                     % Do not remove
\institute{National Centre for Nuclear Research, ul.\ Andrzeja So\l tana 7, 05-400 Otwock, Poland
\and Department of Physics, Florida State University, Tallahassee, Florida 32306, USA
\and Heavy Ion Laboratory, University of Warsaw, ul.\ Pasteura 5a, 02-093 Warsaw, Poland}
\date{Received: date / Revised version: date}
\abstract{We discuss the impact of the uncertainty ($\pm 8$ keV) in the excitation energy of the astrophysically
important 6.356 MeV $1/2^+_2$ state of $^{17}$O on the precision with which the Coulomb reduced ANC ($\widetilde{C}$)
for the $\left<^{17}\mathrm{O}(1/2^+_2) \mid \protect{^{13}\mathrm{C}} + \alpha\right>$ overlap 
can be extracted from direct reaction data. We find a linear dependence of $\widetilde{C}^2$ on the binding energy,
the value extracted varying by a factor of 4 over the range $E_{\mathrm{ex}} = 6.356$ -- $6.348$ MeV. This represents
an intrinsic limit on the precision with which $\widetilde{C}^2$ can be determined which cannot be improved
unless or until the uncertainty in $E_{\mathrm{ex}}$ is reduced.
} %end of abstract
\authorrunning{N. Keeley {\em et al.\/}}
\titlerunning{The Coulomb Modified ANC for the \ldots}
\maketitle
During the past decade the so-called asymptotic normalisation coefficient (ANC) \cite{Blo77,Blo84} has been much in vogue to quantify 
the nuclear structure information extracted from analyses of direct reaction data, although the concept dates back
some fifty years \cite{Dos66,But66,Rap68}; there are differences of detail in the definitions of the reduced normalisation 
of Refs.\ \cite{Dos66,But66,Rap68} and the ANC but they are conceptually identical. The ANC has much to recommend it since it
combines in a single number the information contained in the usual spectroscopic factor, $S$, and the bound state radial wave function,
$u(r)$, thus facilitating comparisons between different analyses. However, for transfers of charged particles involving weakly bound states
the usual ANC can become inconveniently large and a Coulomb renormalised ANC, $\widetilde{C}$, was introduced, defined 
as \cite{Muk12}:
\begin{equation}
\widetilde{C} = \frac{\ell !}{\Gamma (\ell + 1 + \eta)} C
\end{equation}
where $\ell$ is the angular momentum of the transferred particle ($a$) relative to the core ($A$), $\eta = Z_a Z_A e^2 \mu_{aA}/k_{aA}$ 
the Sommerfeld parameter, $k_{aA} = \sqrt{2\mu_{aA}\epsilon}$ the wave number, $\mu_{aA}$ the reduced mass, $\epsilon$ the binding
energy and $\Gamma$ the gamma function. $C$ is the usual ANC:
\begin{equation}
C^2 = S \left(\frac{R \; u(R)}{W_{-\eta,\ell+1/2}(2k_{aA}R)}\right)^2
\end{equation}  
for large values of $R$ where the ANC reaches its asymptotic value and $W$ is the Whittaker function of the second kind.
Note that the Coulomb renormalised ANC depends on the binding energy $\epsilon$ through both the Whittaker function in the
conventional ANC and the gamma function employed in the renormalisation procedure.

This suggests the possibility that for states close to threshold any uncertainty in the
excitation energy $E_{\mathrm{ex}}$ and hence the binding energy $\epsilon$ may have an impact on the accuracy with which
the ANC can be determined from fits to reaction data. One such state is the $^{17}$O $1/2^+_2$
with an excitation energy of 6.356 MeV, $\sim 3$ keV below the $\alpha$ emission threshold, and a stated
uncertainty of $\pm 8$ keV \cite{Til93}. 
The ANC for the $\left<^{17}\mathrm{O}(1/2^+_2) \mid\protect{^{13}\mathrm{C}} + \alpha\right>$ overlap 
is astrophysically important since the $^{13}$C($\alpha$,$n$)$^{16}$O reaction is considered to be
the main source of neutrons for the s process in asymptotic giant branch (AGB) stars \cite{Ibe83}.
The reaction rate at the energies required has to be extrapolated from higher energy data and the presence
of the 6.356 MeV $1/2^+_2$ state in $^{17}$O complicates matters since it enhances the low energy cross
section, making an important contribution to the astrophysical $S$ factor. An accurate determination of
the Coulomb reduced ANC for the $\left<^{17}\mathrm{O}(1/2^+_2) \mid\protect{^{13}\mathrm{C}} + \alpha\right>$ overlap
is required to fix this contribution and there is a considerable literature on the subject, see e.g.\ Ref.\
\cite{Avi15} and references therein. 

To test the possible sensitivity of $\widetilde{C}^2$ to the binding energy we re-analysed two typical data sets for transfer reactions
probing the $\left<^{17}\mathrm{O}(1/2^+_2) \mid\protect{^{13}\mathrm{C}} + \alpha\right>$ overlap, the 3.57 MeV 
$^{13}$C($^6$Li,$d$)$^{17}$O data of Ref.\ \cite{Avi15} and the 45 MeV $^{13}$C($^{11}$B,$^7$Li)$^{17}$O data of Ref.\ \cite{Mez17}. 
It was assumed that the distorted wave Born approximation (DWBA) is adequate to describe the reaction process in both cases (the $^{13}$C($^{11}$B,$^7$Li)$^{17}$O
calculations are technically coupled channels Born approximation (CCBA) since they included coupling to the first excited state of $^{11}$B, but only the direct transfer
step from the ground state of $^{11}$B was included) and most inputs---distorting potentials and projectile overlaps---were retained 
from the original publications.

For the target overlap, the transferred $\alpha$ particle was bound to the $^{13}$C core in a conventional Woods-Saxon well with
radius and diffuseness parameters $R = r_0 \times 13^{1/3}$ fm, $a_0 = 0.65$ fm. The value of $r_0$ was varied from $1.25$ to
$2.50$ in steps of $0.05$ and for each value calculations were performed varying the value of $\epsilon$ from the nominal value of
$2.69$ keV to $10.69$ keV, corresponding to the lower limit of the uncertainty in the excitation energy. Note that the effect of
decreasing $\epsilon$ was not investigated since this would lead to the state becoming unbound with respect to $\alpha$ emission
and thus make extraction of an ANC problematical, since the bound-state wave function no longer decays 
as a smooth exponential as a function of radius in that case. To avoid subjective
judgements as far as possible the calculations were normalised to the data by minimising $\chi^2$. All calculations were performed
using {\sc Fresco} \cite{Tho88}. 

\begin{figure}
\includegraphics[width=\columnwidth,clip=]{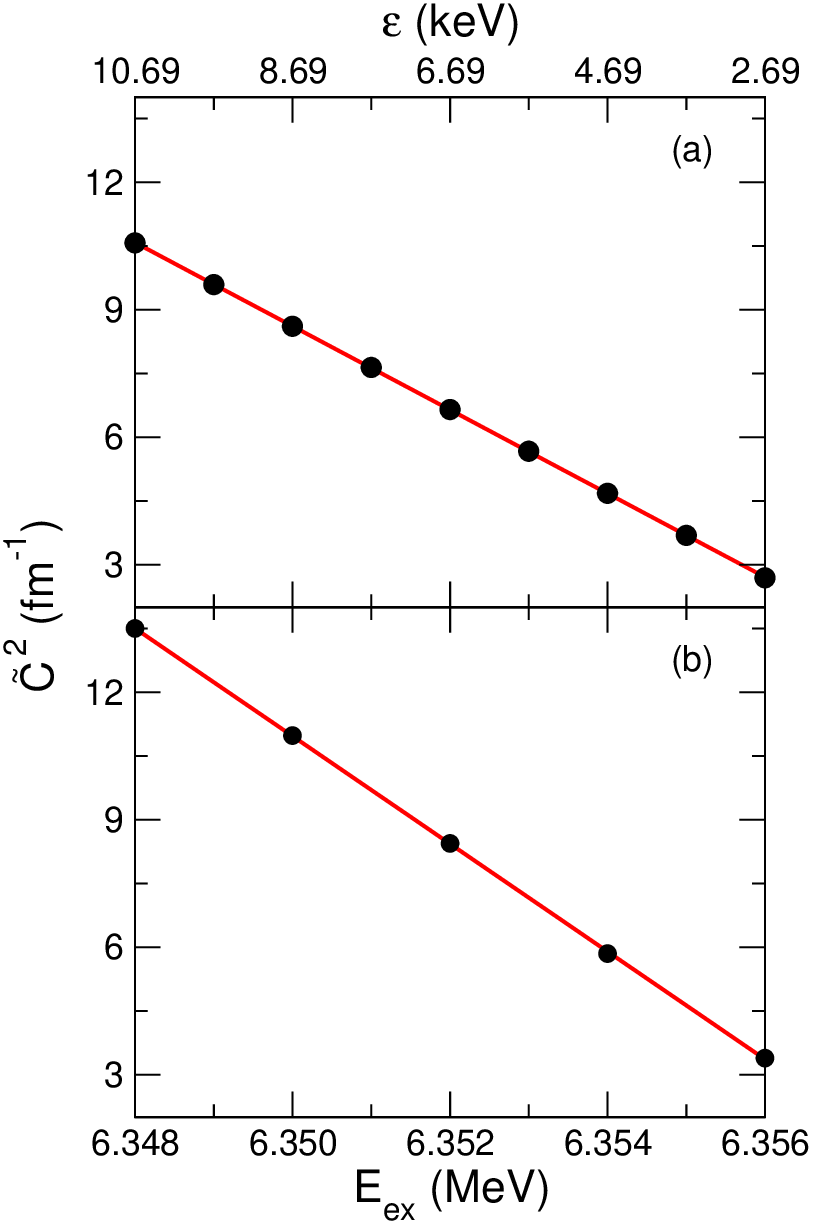}
\caption{\label{fig1}$\widetilde{C}^2$ for the $\left<^{17}\mathrm{O}(1/2^+_2) \mid
\protect{^{13}\mathrm{C}} + \alpha\right>$ overlap as a function of the excitation energy, $E_{\mathrm{ex}}$, and the
binding energy of the transferred $\alpha$ particle, $\epsilon$, extracted from (a) the 
$^{13}$C($^6$Li,$d$)$^{17}$O data of Ref.\ \cite{Avi15} and (b) the $^{13}$C($^{11}$B,$^7$Li)$^{17}$O data of Ref.\ \cite{Mez17}.
The lines represent straight line regression fits to the values obtained (filled circles).} 
\end{figure}

For a given value of $r_0$ the calculated angular distributions and the associated spectroscopic factors do not vary as a function of
$\epsilon$ over the range tested here. However, the Coulomb renormalised ANC varies considerably, see Fig.\ \ref{fig1}.
The values of $\widetilde{C}^2$ plotted in Fig.\ \ref{fig1} are for binding potentials with $r_0 = 1.50$ but similar results
were obtained for all $r_0$. The square of the Coulomb renormalised ANC increases by approximately a factor of 4 as $\epsilon$
is increased from $2.69$ keV to $10.69$ keV. Furthermore, the variation of $\widetilde{C}^2$ as a function of $\epsilon$ is
linear, the solid lines in Fig.\ \ref{fig1} represent linear regression fits to the individual values of $\widetilde{C}^2$
denoted by the filled circles. We emphasise that the different $\widetilde{C}^2$ were obtained from absolutely identical fits
to the data---the calculated angular distributions are graphically indistinguishable---and that for a given $r_0$ the extracted
spectroscopic factors do not vary as a function of $\epsilon$ over this range. The slight offset in the absolute values of
$\widetilde{C}^2$ extracted from the two reactions is not significant; since it is the product of the projectile and
target overlap ANCs that is actually determined by the fit to the data the choice of projectile overlap ANC will obviously
affect the absolute value of the target overlap ANC.

In summary, we find that due to the uncertainty of $\pm 8$ keV in the excitation energy of the near threshold 6.356 MeV
$1/2^+_2$ state of $^{17}$O there is currently an intrinsic limit of approximately a factor of 4 in the precision with
which the Coulomb modified ANC for the $\left<^{17}\mathrm{O}(1/2^+_2) \mid \protect{^{13}\mathrm{C}} + \alpha\right>$ overlap
can be determined, with all that this entails for the astrophysical $S$ factor for the $^{13}$C($\alpha$,$n$)$^{16}$O reaction. 
This will not be improved until or unless the uncertainty in the excitation energy for this state is
reduced. It may well be that this is a unique, not to say pathological, case since the 6.356 MeV $1/2^+_2$ state is
above the neutron emission threshold and at the same time very close to the $\alpha$ emission one. Nevertheless, our
results show that for near-threshold levels the uncertainty in the value of the excitation energy may have a significant
impact on the ANC and this should be tested on a case-by-case basis. We note that uncertainties in the masses
of the core (A) and/or composite (B) nuclei may also impact the precision with which the ANC for the $\left<\mathrm{B}
\mid \mathrm{A} + \mathrm{a}\right>$ overlap can be determined via the consequent uncertainty in the 
$\mathrm{B} \rightarrow \mathrm{A} + \mathrm{a}$ separation energy when either or both are exotic nuclides. Such an effect has been
noted by Ogata \cite{Oga12} in the context of Eikonal reaction theory.


\begin{thebibliography}{}
\bibitem{Blo77}L. D. Blokhintsev, I. Borbely, and E. I. Dolinskii, Fiz.\ Elem.\ Chastits At.\ Yadra {\bf 8}, 1189 
(1977) [Sov.\ J. Part.\ Nuclei {\bf 8}, 485 (1977)].
\bibitem{Blo84}L. D. Blokhintsev, A. M. Mukhamedzhanov, and A. N. Safronov, Fiz.\ Elem.\ Chastits At.\ Yadra {\bf 15}, 
1296 (1984) [Sov.\ J. Part.\ Nuclei {\bf 15}, 580 (1984)].
\bibitem{Dos66}M. Dost and W. R. Hering, Z. Naturforsh.\ {\bf 21A}, 1015 (1966).
\bibitem{But66}P. J. A. Buttle and L. J. B. Goldfarb, Nucl.\ Phys.\ {\bf 78}, 409 (1966).
\bibitem{Rap68}J. Rapaport and A. K. Kerman, Nucl.\ Phys.\  A {\bf 119}, 641 (1968).
\bibitem{Muk12}A. M. Mukhamedzhanov, Phys.\ Rev.\ C {\bf 86}, 044615 (2012).
\bibitem{Til93}D. R. Tilley, H. R. Weller, and C. M. Cheves, Nucl.\ Phys.\ A {\bf 564}, 1 (1993). 
\bibitem{Ibe83}I. Iben and A. Renzini, Annu.\ Rev.\ Astron.\ Astrophys.\ {\bf 21}, 271 (1983).
\bibitem{Avi15}M. L. Avila, G. V. Rogachev, E. Koshchiy, L. T. Baby, J. Berlarge, K. W. Kemper, A. N. Kuchera, and D. Santiago-Gonzalez,
Phys.\ Rev.\ C {\bf 91}, 048801 (2015).
\bibitem{Mez17}S. Yu.\ Mezhevych, A. T. Rudchik, A. A. Rudchik, O. A. Ponkratenko, N. Keeley, K. W. Kemper, M.~Mazzocco, K. Rusek, and
S. B. Sakuta, Phys.\ Rev.\ C {\bf 95}, 034607 (2017). 
\bibitem{Tho88}I. J. Thompson, Comput.\ Phys.\ Rep.\ {\bf 7}, 167 (1988).
\bibitem{Oga12}K. Ogata, Prog.\ Theor.\ Phys.\ Suppl.\ {\bf 196}, 203 (2012).
\end{thebibliography}
\end{document}